\documentclass[conference]{IEEEtran}

\usepackage[ruled]{algorithm2e}

\usepackage{tikz}
\usepackage{pgf}
\usepackage{amssymb}
\usetikzlibrary{arrows,positioning,calc,fit,backgrounds,shapes,automata,,decorations,tikzmark}
\usepackage{pgfplots,pgfplotstable} 
\usepgfplotslibrary{units}
\usetikzlibrary{external} 
\usetikzlibrary{spy,shapes.misc,shapes.geometric}
\usepgfplotslibrary{groupplots,external}

\usepackage{enumitem}

\newcolumntype{M}[1]{>{\centering\arraybackslash}m{#1}}

\usepackage{array}
\usepackage[thinlines]{easytable}
\usepackage{multirow}
\usepackage{listings}
\usepackage{xspace}

\usepackage{mathrsfs}
\usepackage{filecontents}

\usepackage{amsmath}
\usepackage[skins,theorems]{tcolorbox}
\tcbset{highlight math style={enhanced,
		colframe=black,colback=white,arc=5pt,boxrule=0.7pt}, width=0.95\columnwidth}
	
\usepackage{amsfonts}
\usepackage{amssymb}
\usepackage{calligra}
\usepackage{calrsfs}

\usepackage[mathscr]{euscript}

\newtheorem{lemma}{Lemma}

\usepackage{acronym}

\newcommand{\conSym}{\mathcal{C}}

\newcommand{\TCSee}{$\Delta_{LS_0 \to E_1}$}
\newcommand{\mTPCsamp}{$T_{PCsamp}$}  
\newcommand{\mTPCproc}{$T_{PCproc}$}

\newcommand{\realN}{ \mathbb{R}\xspace} 
\newcommand{\noAssumptions}{ \top\xspace } 
\newcommand{\within}{\xspace\textbf{within}\xspace} 
\newcommand{\every}{\xspace\textbf{every}\xspace} 

\newcommand{\ignore}[1]{{}}
\newcommand{\codeFont}[1]{\texttt{#1}}

\definecolor{graphFirst}{RGB}{2,136,209} 
\definecolor{graphSecond}{RGB}{211,47,47} 
\definecolor{graphThird}{RGB}{245,124,0} 
\definecolor{graphFourth}{RGB}{56,142,60} 
\definecolor{graphFifth}{RGB}{81,45,168} 
\definecolor{graphSixth}{RGB}{69,90,100} 
\definecolor{graphSeventh}{RGB}{251,192,45} 

\definecolor{backgroundFirst}{RGB}{129,212,250} 
\definecolor{backgroundSecond}{RGB}{239,154,154} 
\definecolor{backgroundThird}{RGB}{255,204,128} 
\definecolor{backgroundFourth}{RGB}{165,214,167} 
\definecolor{backgroundFifth}{RGB}{179,157,219} 
\definecolor{backgroundSixth}{RGB}{176,190,197} 
\definecolor{backgroundSeventh}{RGB}{255,245,157} 

\usepackage{xargs}   
\usepackage[colorinlistoftodos,prependcaption,textsize=tiny]{todonotes}
\newcommandx{\unsure}[2][1=]{\todo[linecolor=red,backgroundcolor=red!25,bordercolor=red,#1]{#2}}
\newcommandx{\change}[2][1=]{\todo[linecolor=blue,backgroundcolor=blue!25,bordercolor=blue,#1]{#2}}
\newcommandx{\info}[2][1=]{\todo[linecolor=OliveGreen,backgroundcolor=OliveGreen!25,bordercolor=OliveGreen,#1]{#2}}
\newcommandx{\improvement}[2][1=]{\todo[linecolor=Plum,backgroundcolor=Plum!25,bordercolor=Plum,#1]{#2}}
\newcommandx{\thiswillnotshow}[2][1=]{\todo[disable,#1]{#2}}

\begin{document}
	
	
	\acrodef{ALM}{Adaptive Logic Module}
\acrodef{APD}{Action Potential Duration}
\acrodef{BCL}{Base Cycle Length}
\acrodef{CPS}{Cyber-Physical System}
\acrodef{iCPS}{Industrial Cyber-Physical System}
\acrodef{DDS}{Data Distribution Service}
\acrodef{DI}{Diastolic Interval}
\acrodef{DSP}{Digital Signal Processor}
\acrodef{DTTS}{Discrete Time Transition System}
\acrodef{DHA}{Deterministic Hybrid Automata}
\acrodef{EA}{Evolutionary Algorithm}
\acrodef{ECG}{Electrocardiogram}
\acrodef{EGM}{Electrogram}
\acrodef{FPGA}{Field Programmable Gate Arrays}
\acrodef{HA}{Hybrid Automata}
\acrodef{HIOA}{Hybrid Input Output Automata}
\acrodef{ILP}{Integer Linear Programming}
\acrodef{MCU}{Microcontroller Unit}
\acrodef{ODE}{Ordinary Differential Equation}
\acrodef{PoC}{Plant-on-a-Chip}
\acrodef{PLC}{Programmable Logic Controller}
\acrodef{QoS}{Quality of Service}
\acrodef{SHIOA}{Synchronous Hybrid Input Output Automata}
\acrodef{SWA}{Synchronous Witness Automata}
\acrodef{TA}{Timed Automata}
\acrodef{WHA}{Well-formed Hybrid Automata}
\acrodef{WCET}{Worst-Case Execution Time}
\acrodef{FSM}{Finite State Machine}

\acrodef{ST}{Stimulated}
\acrodef{UP}{Upstroke}
\acrodef{ERP}{Effective Refractory Period}
\acrodef{RRP}{Relative Refractory Period}
\acrodef{RP}{Resting Period}
\acrodef{AP}{Action Potential}

\acrodef{SA}{Sinoatrial}
\acrodef{AV}{Atrioventicular}
\acrodef{RVA}{Right Ventricular Apex}

\acrodef{QSS}{Quantized State System}

	
	
	\title{CLAIR: A Contract-based Framework for  Developing   Resilient CPS Architectures}

	\author{
	
	\IEEEauthorblockN{Sidharta Andalam, Daniel  Jun Xian Ng, Arvind Easwaran}
	\IEEEauthorblockA{Nanyang Technological University (NTU), 
		Singapore\\
		Email: arvinde@ntu.edu.sg}
	\and
	\IEEEauthorblockN{Karthikeyan Thangamariappan}
	\IEEEauthorblockA{Delta Electronics Inc.,
		Singapore\\
		Email: karthikeyan.t@deltaww.com}
	
	\thanks{
		This work was conducted within the Delta-NTU Corporate Lab for Cyber-Physical Systems with funding support from Delta Electronics Inc. and the National Research Foundation (NRF) Singapore under the Corp Lab@University Scheme.
	} 
	
}
		
\thanks{
	CLAIR is the French word for clear. It resonates with our approach to improve transparency across layers of a cyber-infrastructure. This research work was conducted within the Delta-NTU Corporate Lab for Cyber-Physical Systems with funding support from Delta Electronics Inc and the National Research Foundation (NRF) Singapore under the Corp Lab@University Scheme.
}

	\maketitle
	
	\begin{abstract}
	Industrial cyber-infrastructure is normally a multi-layered architecture.
The purpose of the layered architecture is to hide complexity and allow independent evolution of the layers.
In this paper, we argue that this traditional strict layering results in \emph{poor transparency}
across layers affecting the ability to significantly improve resiliency.
We propose a contract-based
methodology where components across and within the layers of the cyber-infrastructure
are associated with contracts and a light-weight resilience
manager. This allows the system to detect faults (contract
violation monitored using observers) and react (change contracts
dynamically) effectively. It results in (1)~improving transparency across layers; helps resiliency, (2)~decoupling fault-handling code from application code; helps code maintenance, (3)~systematically generate error-free fault-handling code; reduces development time.
Using an industrial case study, we demonstrate the proposed methodology. 

\end{abstract}
	

	
	
	\section{Introduction}
	\label{sec:introduction}

A key focus of \acf{iCPS} is to
 reduce factory downtime by employing a more intelligent
 distributed system that automatically detects faults
 and dynamically reconfigures to recover from faults 
 without significantly affecting the normal operations~\cite{LEE2015,Vyatkin2017}.
 We refer to such systems as \emph{resilient} systems.

\begin{figure}[htbp]
	\centering
	\includegraphics[width=\columnwidth]{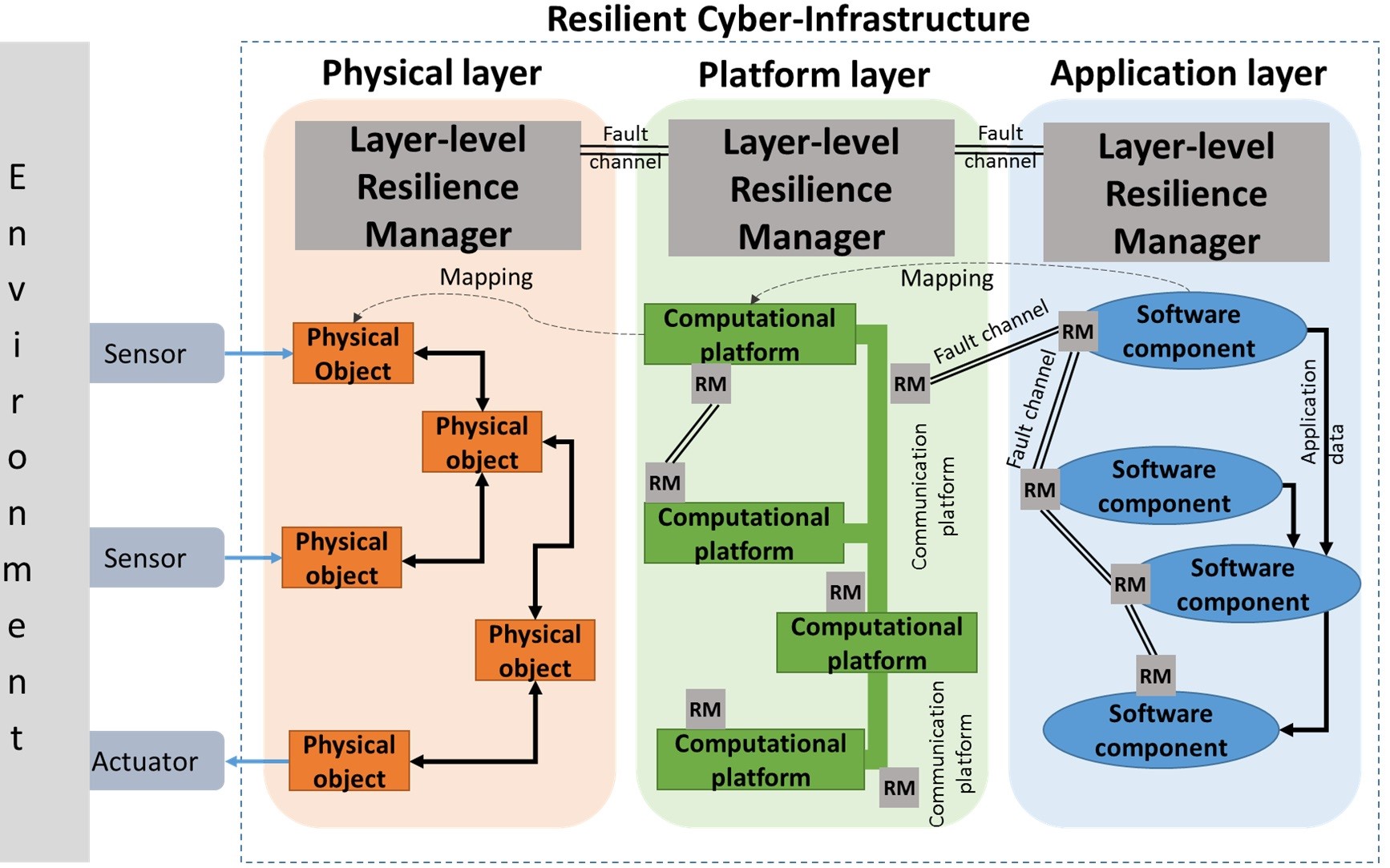}
	\caption{\label{fig:CILayers} Proposed cyber-architecture with integrated resilience manger~(RM) enabling cross-layer component-level interactions.}
\end{figure}

Existing cyber-architecture for \ac{CPS}  follows the classical layered middleware with three different layers such as application, platform and physical layers~\cite{Sztipanovits2012}, see Figure~\ref{fig:CILayers}. 
(1)~The physical layer comprises physical components such as sensors, actuators, controllers and communication hardware. 
(2)~The platform layer embodies computational and communicational platforms such as operating systems and network managers, respectively.  
(3)~Finally, the application layer accommodates the software components which describe the behaviour of an application.
In this paper, we focus on the following  problems that are applicable across the layers.\\

\noindent\textbf{Problem 1: Lack of  frameworks that enable cross-layer interactions for managing resiliency. } 

{ Traditional hardware based redundancy techniques are expensive. 
In contrast, software based techniques are more affordable and flexible. 
However, they are not very effective due to the lack of 
transparency across the layers of the cyber-infrastructure. 
In the following, we illustrate how cross-layer interactions can improve the quality of resiliency by discussing a few existing resilient architectures. Then, we briefly discuss our framework which enables cross-layer interactions to manage resiliency.
}

In RIAPS~\cite{Eisele2017} architecture, the focus is on developing a distributed resilient CPS. As an example, a resilient discovery service~(DS)  was explored.
 Using heartbeat signals and timestamps, DS detects a failure of a publisher/subscriber pair. When  a failure occurs, DS de-registers the pair from the list of registered services.  A publisher/subscriber needs to re-register once they become active. The process of de-registering  or re-registering  is very time consuming and it also needs to be communicated with neighbouring nodes. 
 In this scenario, the cause of the failure (e.g., intermittent fault in the physical layer) and the expected recovery time (e.g. 2 seconds or 2 hours)  were not available to the discovery service manager (residing in the platform layer).
 If the recovery time information was available to the 
 DS manager, it can choose not to de-register a publisher/subscriber and avoid unnecessary time consuming registration process across all neighbouring nodes.

In iLand~\cite{Valls2013}  architecture, the focus is on developing a reconfigurable service oriented distributed system. An application is described as a graph, where each  vertex is a service provided by a component of the system. Interestingly, each service may have zero or more alternative services. 
At runtime, the reconfiguration manger may select an alternative service based on faults. Once again, the reconfiguration manager that is residing in the platform layer is unaware of the fault type or the recovery time. 
If the platform layer could provide the manager with the list of services that will be affected due to
the fault, the manager can select the right order of services in the application graph.
This reduces the number of reconfigurations. 
The paper does not focus on the mechanisms for detecting faults.
Also, the architecture design does not discuss any cross-layer interactions to improve resiliency.

For industrial applications based on the IEC~61499 standard~\cite{IEC61499s1,IEC61499s2,IEC61499s3}, 
an approach to resiliency is provided using a runtime reconfiguration manager~\cite{Vyatkin2017}. Failure of node~$n$ executing a publisher~$p$ is detected by the manger using heartbeats. Then, based on ontology~\cite{Vyatkin2017,OliverH2017}, node~$n^\prime$ is identified as an alternative to host the new producer  $p^\prime$. 
The new producer is created dynamically and
 the connection to consumer is modified.  
 Once again, the manager residing in the application layer is unaware of the 
 cause of the node failure. 
 If the platform layer could provide the manager with an
 expected recovery time, the manager could avoid a time consuming reconfiguration process.


Overall, we believe that there is a \emph{need to improve transparency across layers such that they work together to satisfy resiliency needs effectively and at low cost.}






\noindent\textbf{ }






\noindent\textbf{Problem 2: Code pollution due to intertwined application code and fault handling code. } 
Managing the increasing functional and safety requirements of industrial automation systems 
is a daunting process. It results in large amount of code that are hard to understand. 
The problem is elevated even  further as application code is intertwined with fault handling code due to lack of 
clear guidelines~\cite{Steinegger2013}.
In typical manufacturing applications,   
fault handling code on average takes the lion's share amounting to nearly 
83\%~\cite{Steinegger2013,Gttel2013}. 
Most of this code is manually written and hence error prone, making it harder to certify.

For industrial controllers, a fault isolation methodology is presented using diagnostics automaton that can observe an order/pattern of events that leads to a fault~\cite{Luca20111}. 
This approach is prone to the well-known state-explosion problem.
Also, it is not expressive enough to describe fault detection techniques based on dynamic models such as \acfp{ODE}~\cite{Isermann2005}. 

 Overall, we believe that there is a \emph{need for formal approach that systematically decouples fault-handling techniques from application code, and automatically generate fault-handling code with minimal user intervention.}\\

\textbf{Proposed solution.} In our approach, the application and the platform layers are described as  a network of components.
Components 
across and within the layers of the cyber-infrastructure
are associated with a  resilience manger that ensures 
lightweight fault monitoring and response. 
We use formal contracts to capture the assumptions on the behaviour of the environment and guarantees about the behaviour of the component~\cite{benveniste2015contracts}.
A failure of a contract is treated as a fault.
In response to a fault, resilience manager communicates
with the resilience managers of other components to find a feasible solution efficiently.
E.g., if an object detection component on a conveyor belt of an assembly line is unable to meet its deadline (failure of time-based contract), as a response the resilience manger can switch to a less time consuming detection algorithm or communicate with the motor to temporarily reduce the speed of the conveyor belt.

Overall, the resiliency of the system is managed by two types of resilience managers.
(a)~A \emph{component-level} resilience manager (denoted as RM in Figure~\ref{fig:CILayers}) that is associated to a single component.
(b)~A \emph{layer-level} resilience manager (see top of the figure) that is associated to each layer.  
For the case when a component-level resilience manager is unable to find a feasible solution, it informs the layer-level resilience manger. 
Our intuition is that at runtime, most of the resiliency issues can be handled
by the  component-level resilience manager.
Finally, fault channels are used for communication between resilience managers. 
The proposed methodology was briefly introduced earlier~\cite{andalam2018}. In comparison, this paper presents a more detailed architecture and implementation and validation of the proposed methodology.



The proposed contract-based methodology (CLAIR) for resiliency can be integrated with existing 
frameworks for designing cyber-infrastructure for CPS such as METROII~\cite{Davare2013}, OpenMETA~\cite{Sztipanovits2014},  RIAPS~\cite{Eisele2017} and iLAND~\cite{Valls2013}.
They present complementary features to manage  application, resources, devices, logs and security.


\textbf{Key contributions of the paper:} 
\begin{enumerate}
	\item We propose a new contract-based methodology to enable resilient cyber-architecture. Components  across and within the layers of the architecture
	are associated with contracts and a light-weight resilience manager, allowing cross-layer interactions.
	\item  We use contracts to formally capture the requirements of a component. This allows us to systematically generate observers what are based on computational models such as \ac{FSM}, \ac{TA} and \ac{HA}. These observers are independent of the application/component behaviour, avoiding code pollution. 
	\item We implement the architecture using industry standards such as  IEC~61499 and \ac{DDS}. Using an industrial application, we illustrate the features of the proposed contract-based methodology.
\end{enumerate}



	\section{The proposed framework}
	\label{sec:archReq}

In the following we discuss some of the non-functional requirements of the proposed architecture and our approach towards implementing them.

\textbf{1. Cross-layer Interactivity}: Allows fine-grained
cross-layer communication such that a fault detection and its handling can be implemented across various layers, resulting in a cost-effective and robust cyber-infrastructure.
In our approach,  component-level resilience manger interacts with components within and across layers. This helps to reduce fault detection and handling time. 
Furthermore, the layer-level resilience manager provides a more centralised solution for resiliency.

\textbf{2. Composability}:  We follow the well know component-based design methodology
to ensure functional properties of a component are not influenced by other components. Component behaviour is not altered due to the interactions with other components. Furthermore, system-level properties can be realised from a network of component-level properties~\cite{Sztipanovits2012}.
However, non-functional requirements such as timing are hard to guarantee as multiple components of an application layer can be mapped to a single component of a platform layer.

\textbf{3. Dynamicity}:
 Components are added/removed dynamically. The system should be flexible, self-aware and self-optimise based on  the availability of the resources~\cite{Vyatkin2017}. In our approach, we  depend on existing discovery services~\cite{Eisele2017,DDSrti} and the resilience manager to marshal additional resources. 

\textbf{4. Adaptation Quality}: Given the limited shared resources and possible conflicting recovery strategies, the system should reason about the quality of adaptations to disturbances~\cite{Denker2012}. In future, we plan to incorporate a multi-dimensional resilience metric~\cite{Friedberg2016} to improve resiliency. 
The metric needs to be  abstract and integrates information across the layers.
Also, we need to understand the impact of a failure at component-level on the system-level properties~\cite{GOSSLER2015}.

\ignore{
\begin{table}[htbp]
	\centering
	\caption{\label{tab:archReq}Non-functional requirement of the  cyber-infrastructure}
	\begin{tabular}{|m{0.30\columnwidth}|m{0.65\columnwidth}|}\hline
		\textbf{\small Requirements }
		& \textbf{\small Our Solutions}
		\\\hline
		{1. Cross-layer Interactivity}  
		& 
		Component-level  resilience manger for localised fault handling and
		layer-level resilience manager a more optimised solution.
		\\\hline
		{2. Composability}
		&  Component-based design with contracts for interface~\cite{Sztipanovits2012}.
		\\\hline
		{3. Dynamicity}
		& Based on existing discovery services~\cite{Eisele2017} (not within the scope of this paper).
		\\\hline
		{4. Adaptation Quality}
		& Multi-dimensional resilience metric~\cite{Friedberg2016}.  Understanding the impact of a failures~\cite{GOSSLER2015} (not within the scope of this paper).
		
		\\\hline
	\end{tabular}
\end{table}
}%

\subsection{Overview of a component}
\label{sec:componetOverview}

\begin{figure}[htbp]
	\centering
	\includegraphics[width=0.85\columnwidth]
	{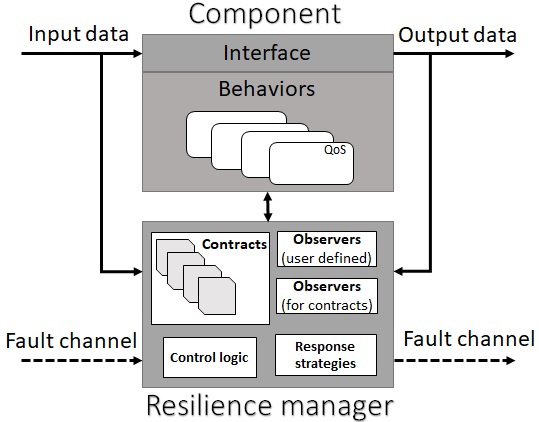}
	\caption{\label{fig:CompArch} Architecture overview of a component and a resilience manager.
	}
\end{figure}

Figure~\ref{fig:CompArch} presents an overview of the component and the resilience manager.
A component is an open system that (1)~receives inputs from the environment,
(2)~executes a behaviour, and (3)~generates output to the environment. 
The environment could be the collection of other components or the physical world.

\begin{itemize}[leftmargin=*]
	\item \textbf{Interface:}
	It defines the Input/Output data channels of a component. 
	
	\item \textbf{Behaviours:} Multiple behaviours can be defined for a given interface.   
	The resilience manager dynamically selects the behaviour of the component based on requirements.

	\item \textbf{Contracts:} It clearly captures the assumptions on the behaviour of the environment, and guarantees about the behaviour of the component~\cite{benveniste2015contracts}. 
	At runtime, the resilience manager can switch between contracts to react to the disturbances in the system.
	
	\item \textbf{Observers}: It monitors the system requirements at runtime~\cite{Ding2008,Bhatti2011,Mhamdi2016}. We express them using   formal models such as finite state machine~\cite{Bhatti2011}, timed automaton~\cite{Alur:1994,Mhamdi2016} or  hybrid automaton~\cite{Henzinger1996}.
	
	\item \textbf{Resilience manager:} Detects faults (using observers) and decides (control logic) the best course of action. It also responds to  fault information from other components, via fault channels.

\end{itemize}

\subsection{Overview of the design flow}
\begin{figure}[htbp]
	\centering
	\includegraphics[width=\columnwidth]{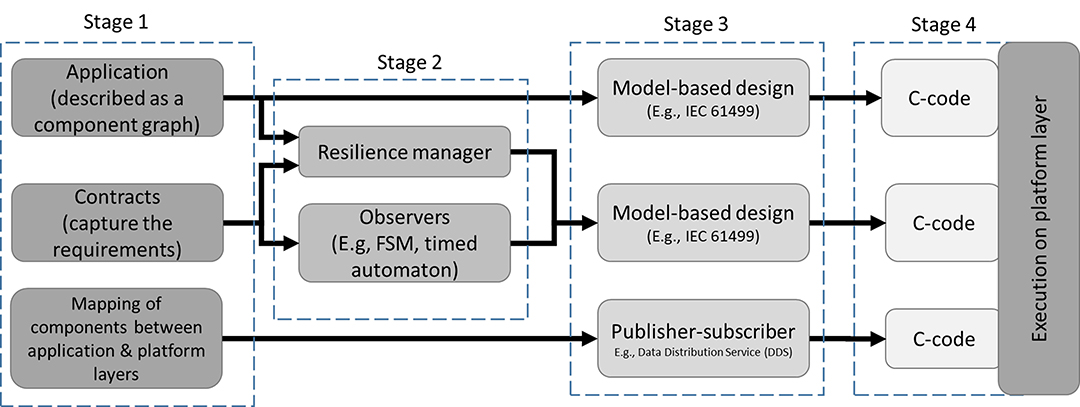}
	\caption{\label{fig:flow} Overview of the design flow for the proposed methodology.}
\end{figure}

Figure~\ref{fig:flow} presents an overview of the design flow.
In stage~1, (a) an application is described as a component graph,
 (b) mapping between the components of the application and platform layer is provided and (c) the requirements are captured formally using contracts~\cite{benveniste2015contracts}.
In stage~2, observers are generated based on contract specifications.
 They check the validity of the contracts at runtime. 
In stage~3, we generate computation and communication models using industry standard such as IEC~61499 and \ac{DDS}. 
Finally, in stage~4, using off-the-shelf tools we generate C-code which is executed by the platform layer. For each stage, we now elaborate on our design choices and their advantages.

\textbf{Stage~1.} We have chosen to describe the application as a component graph because component-based software engineering  has been successfully used for large-scale system designs. 
It relies on the concept of developing basic reusable components  with well defined interfaces.  
During integration of components, their well-defined interfaces ensure easy assembly.  They are used in popular software tools such as Simulink~\cite{SimulinkTool}.  

Requirements engineering presents a major challenge
for software development. Poorly managed and ill-defined requirements
lead to lack of visibility into changing requirements  and  hinders traceability between requirements and implementation. Using contracts~~\cite{benveniste2015contracts}, 
 we (1)~clearly define the requirements, (2)~improve visibility due to concise and formal descriptions. More importantly, contracts allow us to systematically decouple fault-detection code from the application code.

\textbf{Stage~2.}
Static verification techniques are not generally adequate to validate whether or not
the contracts are satisfied. This may be because
some of the requirements can only be verified with the data available at runtime
(e.g., a sensor producing invalid data). As an alternative,  we use \emph{observers}
to monitor the contracts at runtime~\cite{Ding2008,Mhamdi2016}. To observe static and dynamic behaviour of a system, we express observers using computational models such as finite state machine~\cite{Bhatti2011}, timed automaton~\cite{Alur:1994,Mhamdi2016} and  hybrid automaton~\cite{Henzinger1996}.
Furthermore, due to the well-defined computation models, 
the executable code (fault-handling code) can be automatically generated  with minimal human intervention. This ensures error-free production ready code.

\textbf{Stage~3.} Since we are targeting industrial automation, we have chosen to implement the application based on the IEC~61499 standard~\cite{IEC61499s1,IEC61499s2,IEC61499s3}. It uses the component-based engineering to improve software quality and reduce the development time. Importantly, the standard provides a portable high-level executable specification framework for distributed automation. It also allows us to develop reconfigurable applications enabling self-adaptive cyber-physical systems~\cite{Vyatkin2017}.

For communication between the components, we have chosen to use
\ac{DDS} as it enables communication mechanisms that go beyond the classic publish-subscribe model~\cite{Pardo2003,DDSrti}. 
 It can handle the interruptions when a publisher/subscriber  is temporarily or permanently unavailable. Furthermore,  it allows us to specify QoS parameters over the communication between a publisher and a subscriber.

\textbf{Stage~4.} Finally, we have chosen C~language because it is widely supported by many micro-controllers.


\section{Example application}

\begin{figure}[thbp]
	\centering
	\includegraphics[width=0.99\columnwidth]{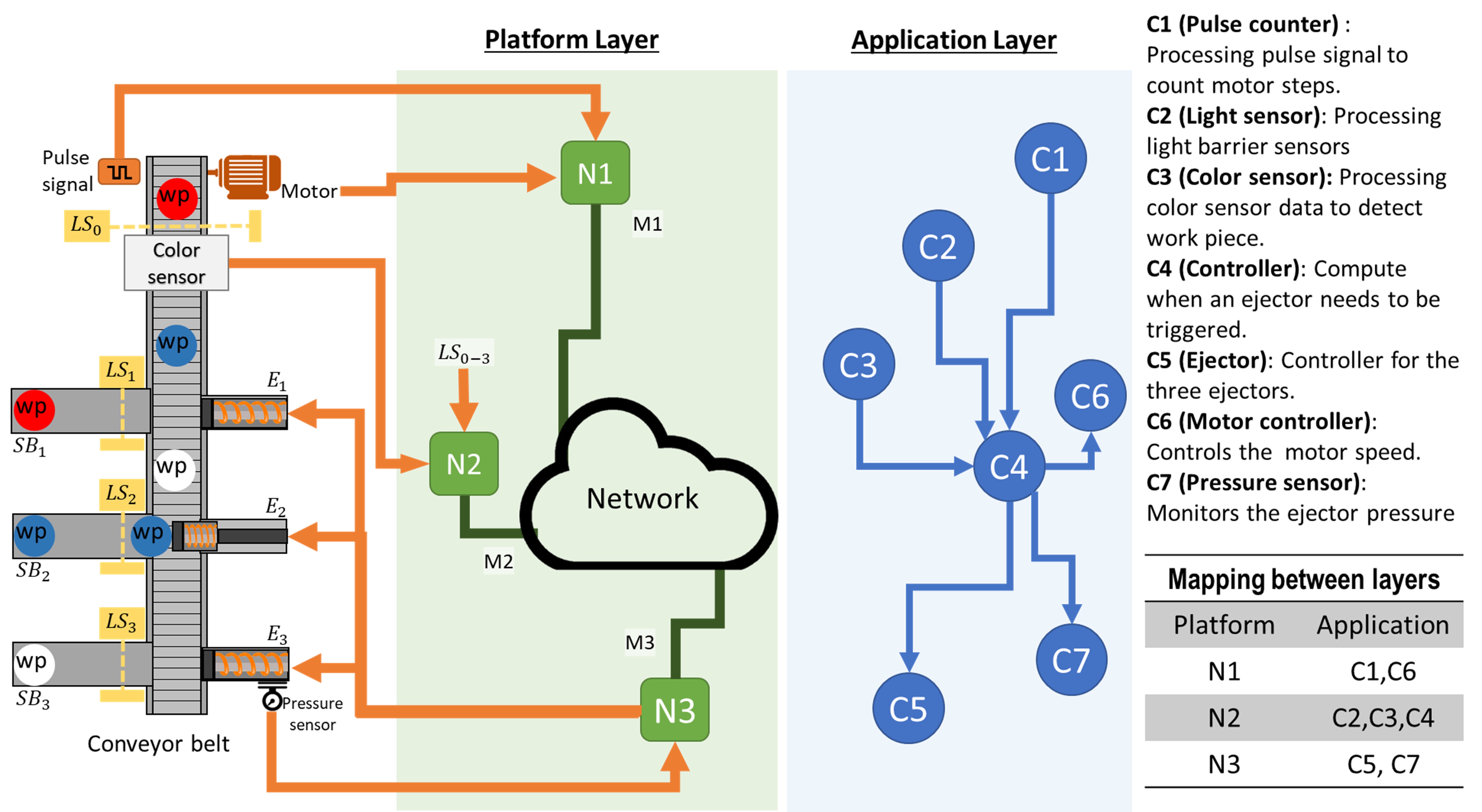}
	\caption{\label{fig:CIexample} An industrial assembly line sorting application. A colour sensor is used to detect the colour of the work pieces (WP) on a continuously moving conveyor belt and ejectors ($E_1, E_2, E_3$) are used push the work pieces into sorting bins ($SB_1, SB_2, SB_3$). Light sensors ($LS$) are used as barriers to detect a work piece. The figure shows a blue work piece being pushed into bin $SB_2$ by the ejector~$E_2$. The application is described as a network of components ($C_1,\dots,C_7$) that are mapped to  the components in the platform layer  ($N_1,\dots,N_3$) . 
	}
\end{figure}

Figure~\ref{fig:CIexample} presents the running example of this paper which reflects a typical assembly line setup in manufacturing.
The goal of the application
is to successfully identify the work pieces (WP) by their color (red, blue and white) and sort them into their respective storage bins ($SB_{1}$, $SB_{2}$ and $SB_{3}$). The application relies on input from a color sensor (to detect the color of the WP),  pulse signal from an encoder (for computing the motor steps), and 3 light sensors to detect the position of a work piece. Also, the application controls the actuators such as the three ejectors ($E_1$, $E_2$ and $E_3$) and a motor.

Given the assembly line is continuously moving and the  cyber-infrastructure is distributed, 
the challenge  is to 
process sensor information and to activate the ejectors at
the right time such that the work pieces reach their respective storage bins. 
Furthermore, the system needs to be resilient to disturbances in the cyber-infrastructure. 

\subsection{Application graph}
The assembly line application is  described as an  \emph{application graph}. The nodes  and edges of the graph represent components
and  communication between the components, respectively.
Figure~\ref{fig:CIexample} shows the application graph with  seven components $C_1,\dots, C_7$.
Component $C_1$ periodically samples the pulse signal to count the number of motor steps. The four light barriers ($LS_0,\dots, LS_3$) are sampled periodically by Component~$C_2$. Component $C_3$ periodically samples from the colour sensor ($CS$).
Component~$C_4$ computes when an ejector needs to be triggered based on information form the sensors. The three ejectors ($E_1, E_2, E_3$) are controlled via components $C_5$. Component $C_6$ controls the speed of the motor (and the conveyor belt). Finally, $C_7$ periodically measures the pressure to detect
leaks in the air pressure controlled ejectors.


\subsection{Mapping from application graph to platform graph}
The platform layer comprises of computational platforms
($N_1,\dots, N_3$) and connecting communication platforms
($M_1, \dots, M_3$). The mapping of the components in an application
graph to computational platforms is shown in Figure~\ref{fig:CIexample}.
An example of a computation platform is a Linux OS and a communication platform is a software defined network.

Earlier in Section~\ref{sec:componetOverview}, we described that a component may have multiple behaviours. E.g., component~$C_1$ is mapped to the computational platform~$N_1$. Assuming that $C_1$ has three possible behaviours ($BEH \in \{beh_1^{c1}, beh_2^{c1}, beh_3^{c1} \}$), based on the platform mapping information we assume that the fixed execution cost can be described using the function $EC:BEH \rightarrow \realN$. 
We also assume that the communication between components is handled by an active network infrastructure. Due to the dynamic nature of the network, the communication cost may not be constant. The cost can be described using the function $CC:BEH \rightarrow \realN$. 

\subsection{Application requirements}
\label{sec:appReq}
Using the constant speed of the conveyor belt 
we can compute the time taken for a work piece to travel from  the first
light sensor ($LS_0$) to an actuator like ejector $E_1$, 
denoted as \TCSee.  
We assume that the decomposition of end-to-end timing constraints (e.g., 4 seconds from $LS_0$ to $E_1$) into  deadlines for each component are given.
E.g., Component~$C_1$ must process the pulse signal within 10ms (\mTPCproc). 

\ignore{

\begin{table}[htbp]
	
	\centering
	\caption{\label{tab:appReq} Some of the requirements for the assembly line application.}
	\begin{tabular}{|m{0.18\columnwidth}|m{0.8\columnwidth}|}\hline
		\textbf{\small Category }
		& \textbf{\small Example requirements}
		\\\hline
		Timing deadline (individual~components)
		& 
		\begin{itemize}[leftmargin=*]
			\item Color sensor has to be sampled at least every  \mTCSsamp~and processed within \mTCSproc.
			\item Pulse sensor (rotary encoder for the motor)  has to be sampled every 	 \mTPSsamp.
			\item Pulse sensor data (binary signal) must be processed to compute the number of motor steps, within \mTPSproc.
		\end{itemize}	
		\\\hline
		Timing deadline (across~components)
		& 
		\begin{itemize}[leftmargin=*]
			\item The conveyor belt is continuously moving at a constant speed. This means the ejectors that push the work pieces have a strict timing constraint. End-to-End delay from the first light sensor~$LS_0$ to each ejector  is as follows.
			\begin{itemize}[leftmargin=*]
				\item For ejector $E_1$ is $T_{LS_0 \to E_1}  (3910~ms)$.
				\item For ejector $E_2$ is $T_{LS_0 \to E_2}  (5290~ms)$.
				\item For ejector $E_3$ is $T_{LS_0 \to E_3}  (6440~ms)$.
			\end{itemize}
		\end{itemize}
		\\\hline
		Functional constraint 
		& 
		\begin{itemize}[leftmargin=*]
			\item The duty cycle of the PWM signal from the motor controller should be within the interval of $[0,100]$.
			\item We consider a pressure leak in the ejectors, when the  pressure from the sensor is different from the expected pressure value by more than $5$~\%. The expected pressure is modelled using a continuous-time model. See Section~\ref{sec:reqDynamic} for more details.
		\end{itemize}
		\\\hline
		Data-driven constraint 
		& 
		\begin{itemize}[leftmargin=*]
			\item Color sensor is based on an infrared sensor. Different environmental parameters such as ambient light and surface of the object influence the sensor readings. Based on experimental data, if the sensor value is within the specified limits, then a work piece is classified as white, red or blue.  Otherwise, they are classified as outliers. See Section~\ref{sec:StochasticContract} for more details.
		\end{itemize}
		\\\hline
		
	\end{tabular}
\end{table}
}


	\section{Contract-based fault-detection}
	\label{sec:designPatt}


In this section, we capture the component-level requirements using contracts based on an existing notation~\cite{benveniste2015contracts}.

\subsection{Timing requirements }
The application component~$C_1$ executing on platform node~$N_1$ is required to 
sample the pulse signal every \mTPCsamp. 
Furthermore, the time required to process the data should be less than \mTPCproc. 
The requirements are captured using the following contract, denoted as symbol $\conSym_{C_1}^{1}$. Symbol~$\noAssumptions$ denotes that the contract does not make any assumptions/constraints.

\begin{equation*}
\tcbhighmath[]{
	\small
	\conSym_{C_1}^{1}:\left\{
	\begin{array}{@{}lll@{}}
	inputs: & \text{$pc$\_data $\in \realN$} \\
	outputs: & \text{$c_1$\_data $\in \realN$} \\
	assumptions: & \text{$\noAssumptions$} \\
	guarantees: &\text{$c_1$\_data} = &\text{process($pc$\_data)}\\
	&&\every~\text{\mTPCsamp}\\
	&&\within~\text{\mTPCproc}
	\end{array}
	\right.
}
\end{equation*}

\ignore{
\subsection{Functional requirements -  Model-based reasoning (static)}
Component~$C_6$ is a motor controller that controls the motor using a PWM signal.
The duty cycle of the signal should always be between zero and one hundred. 
The requirement is captured using the following contract, denoted as symbol $\conSym_{C_6}^{1}$.

\begin{equation*}
\tcbhighmath[]{
	\small
	\conSym_{C_6}^{1}:\left\{
	\begin{array}{@{}lll@{}}
	inputs: & \noAssumptions \\
	outputs: & \text{$c_6$\_data $\in \realN$}  \\
	assumptions: & \text{$\noAssumptions$} \\
	guarantees: &\text{$0 \le c_6$\_data $ \le 100$} 
	\end{array}
	\right.
}
\end{equation*}

\subsection{Functional requirements -  Model-based reasoning (dynamic)}
\label{sec:reqDynamic}
Component~$C_7$ monitors the pressure inside the ejectors.
The expected pressure from the time when the ejector is activated can be modelled using 
${\dot{p}}_{exp}=k_1 \times p_{exp}$ where $p_{exp}(0) = k_2$. 
We consider a pressure leak, when the observed pressure ($p_{obs}(t)$) value from sensor is different from the expected pressure value by more than $5$~\%. 
The requirement is captured using the following contract, denoted as symbol $\conSym_{C_7}^{1}$.

\begin{equation*}
\tcbhighmath[boxsep=0mm, width=1cm]{
	\small
	\conSym_{C_7}^{1}:\left\{
	\begin{array}{@{}lll@{}}
	inputs: & P_{obs} \in \realN \\
	outputs: & isFaulty \in \{0,1\} \\
	assumptions: & \text{$t \ge 0$} \\
	guarantees: & isFaulty = 
		\left\{ 
		\begin{array}{ll}	
		0, & \text{If $0.95 \times P_{exp} \le P_{obs}(t) \le 1.05 \times P_{exp}$}\\
		1, & Otherwise\\
		\end{array}
		\right.
	\end{array}
\right.
	}
\end{equation*}

Due to the continuous-time behaviour of the contract, an hybrid automaton is used to describe the observer, see Section~\ref{sec:obsImp} for more details.

\subsection{Data-driven models}
\label{sec:StochasticContract}

Many industrial processes are complex and  exhibit stochastic behaviour. For such processes, developing accurate models many not be feasible. Thus, many fault detection techniques in industry are mainly data driven~\cite{Chiang2001}.  For example, there are three types of work pieces (white, red and blue). Different environmental parameters such as ambient light and surface of the object influence the sensor readings. Thus, it is hard to develop a physical model. In contrast, data-driven techniques classify  faults based on the collected data, see Figure~~\ref{fig:Shewhart}. The sample consists of 100 work pieces. Each work piece is classified as white, red or blue if the sensor value is within the specified limits. Otherwise, they are classified as outliers.

\begin{figure}[hbtp]
		\begin{tikzpicture}
	\pgfplotsset{width=0.65\columnwidth}

	\begin{axis}[
	enlargelimits=false,
	ymin=500,
	ymax=800,
	ylabel={Optical sensor value},
	xlabel= {Work piece number},
	clip=false,
	]
	\addplot+[
	only marks,
	mark=x,
	color=black,
	mark options={fill=white},
	mark size=2.9pt]
	table[meta=x]
	{./figures/stochastic.dat};
	
	\addplot [gray, no markers] coordinates {(0,525) (100,525)};
	\addplot [gray, no markers] coordinates {(0,558) (100,558)};
	
	\addplot [red, no markers] coordinates {(0,568) (100,568)};
	\addplot [red, no markers] coordinates {(0,590) (100,590)};
	\addplot [blue, no markers] coordinates {(0,750) (100,750)};
	\addplot [blue, no markers] coordinates {(0,755) (100,755)};

	\draw[->, ultra thick] (110,200) -- (90,250) node [pos=0.1,below right] {\small Lower control limit for blue WP};
	
	\draw[->, ultra thick] (110,125) -- (90,95) node [pos=0.1,above right] {\small Upper control limit  for red WP};
	
	\node[text width=3cm] at (125,40) { \color{gray} White work pieces};
	\node[text width=3cm] at (125,75) {\color{red} Red work pieces};
	\node[text width=3cm] at (125,250) {\color{blue} Blue work pieces};
	
	\draw[->, ultra thick] (48,140) -- (42,100) node [pos=0.0, right] {\small Outlier};
	\end{axis}
	\end{tikzpicture}
	\caption{\label{fig:Shewhart} An illustration of the Shewhart chart used in industry~\cite{Chiang2001}. The data reflects the actual color sensor data used by the assembly line sorting example to classify work pieces as blue, red or white.   }
\end{figure}

We propose a stochastic contract based on set theory. 
The control limits (see Fig.~\ref{fig:Shewhart}) for 
blue, red and white work pieces can be represented using intervals
 as $R_{blue}=[750,755]$, 
 $R_{red}=[568,590]$,
  $R_{white}=[535,558]$, respectively.
The requirement to accept the work pieces that only satisfy the control limits is captured using the following contract, denoted as symbol $\conSym_{C_3}^{2}$.

\begin{equation*}
\tcbhighmath[boxsep=0mm, width=1cm]{
	\small
	\conSym_{C_3}^{2}:\left\{
	\begin{array}{@{}lll@{}}
	inputs: & Data_{sensor} \in \realN  \\
	outputs: & Data_{out} \in \realN  \\
	assumptions: & \noAssumptions \\
	guarantees: & Data_{out} \in R_{blue} \cup R_{red} \cup R_{white}  \\
	\end{array}
	\right.
}
\end{equation*}

The Shewchart  uses  a ribbon-like geometrical 
shape to specify the accepted data range. This shape may not always be suitable to capture the requirements.
However, area of an arbitrary shape can  be represented using rectangles which can be mapped to a set of points. 
Of course  there will be a trade-off between precision (identifying outliers) 
and the required computational resource.

}

	
	\label{sec:obsImp}

	\subsection{Observers for runtime validation}
	\label{sec:observers}

The contracts are monitored at run-time using observers. 
In our approach, the observers are expressed using computational models such as finite state machine~\cite{Bhatti2011}, 
timed automaton~\cite{Alur:1994,Mhamdi2016} or  hybrid automaton~\cite{Henzinger1996}.
 In this section, we illustrate that observers can be generated from the contracts, which capture the  requirements.

\ignore{
\subsubsection{\acf{FSM} as an observer}
Finite state machines are dominantly used for fault detection. It captures only discrete timed transitions. For the running example, the contract $\conSym_{C_6}^{1}$ specifies that the value of the  $PWM$ should be between  zero and one hundred. 
Figure~\ref{fig:obsAsFSM} presents the observer that monitors this requirement using a \ac{FSM}. 

\begin{figure}[hbtp]
	\centering
	\begin{tikzpicture}[->,>=stealth',shorten >=1pt,auto,
node distance=4.5cm,
semithick,scale=0.9, transform shape]
\tikzstyle{every state}=[rectangle,rounded corners,
minimum height = 1.5cm, text width=1.2cm, text centered, fill=blue!20,draw=none,text=black, draw,line width=0.3mm]
\tikzstyle{line} = [draw, -latex']

\node[state, fill=green!30]
(PROC)  {Process};

\path[<-, dashed] (PROC.110) edge node[below, align=left, shift={(0.8,1)}] {
	\footnotesize initial \\
	\footnotesize $\begin{matrix}
    {PWM=0}{}
	\end{matrix}$
} ++(0cm,1cm);

\node[state, fill=red!20]
(E1) [right of=PROC] {Error};



\draw (PROC) to[out=0, in=180] (E1) node[ above right =-0.7cm  and 0.2cm of PROC, text width =2cm] {$PWM <~0~or$ $PWM~>~100 $};

\draw (PROC) to[out=200, in=170, distance=1cm ] (PROC) node[ above left = -0.4 and -0.2cm of PROC, text width =2.9cm] {$ 0 \le PWM \le 100 $};

\end{tikzpicture}

	\caption{\label{fig:obsAsFSM} An observer implemented using a finite state machine. It detects if the input PWM signal is outside the accepted range  of $[0,100]$. This observer implements the contract $\conSym_{C_6}^{1}$.  }
\end{figure}

Similarly, a \ac{FSM} with the guard condition of $ Data_{out} \in R_{blue} \cup R_{red} \cup R_{white}$ can be used to implement an observer for the stochastic contract described in Section~\ref{sec:StochasticContract}.
}

\subsubsection{\acf{TA} as an observer}
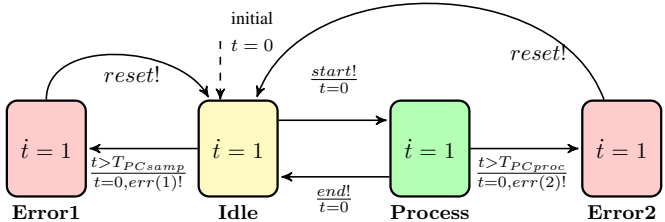
\begin{figure}[hbtp]
	\centering
	\begin{tikzpicture}[->,>=stealth',shorten >=1pt,auto,
node distance=3cm,
semithick,scale=0.85, transform shape]
\tikzstyle{every state}=[rectangle,rounded corners,
minimum height = 1.5cm, text width=1.0cm, text centered, fill=blue!20,draw=none,text=black, draw,line width=0.3mm]
\tikzstyle{line} = [draw, -latex']

\node[state, fill=yellow!30,
label={[shift={(0.0,-2)}]\small $ \mathbf{Idle}$ }]
(IDLE)  {$\dot{t}= 1 $};

\path[<-, dashed] (IDLE.110) edge node[below, align=left, shift={(0.5,1)}] {
	\footnotesize initial \\
	\footnotesize $\begin{matrix}
    {t=0}{}
	\end{matrix}$
} ++(0cm,1cm);

\node[state, fill=green!30,
label={[shift={(0,0)}]}, 
label={[shift={(0,-2)}]\small $\mathbf{Process}$   }]
(PROC) [right of=IDLE] {$\dot{t}=1$};

\node[state, fill=red!20,
label={[shift={(0,0)}]$ $}, 
label={[shift={(0,-2)}]\small $\mathbf{Error 1}$   }]
(E1) [left of=IDLE] {$\dot{t}=1$};

\node[state, fill=red!20,
label={[shift={(0,0)}]$ $}, 
label={[shift={(0,-2)}]\small $\mathbf{Error 2}$   }]
(E2) [right of=PROC] {$\dot{t}=1$};





\path[->] (IDLE.35) edge node[align=center] {
	$\frac{ start! }{t=0}$ \\   
} (PROC.145);

\path[->] (PROC.-145) edge node[align=center] {
	$\frac{ end! }{t=0}$ \\   
} (IDLE.-35);

\path[->] (IDLE) edge node[below, align=left] {
	$\frac{t > T_{PCsamp}}{t=0, err(1)!}$
} (E1);

\path[->] (PROC) edge node[below, align=left] {
	$\frac{t > T_{PCproc}}{t=0, err(2)!}$
} (E2);

\draw (E1) to[out=90, in=120] (IDLE) node[above right =0.2cm of E1] {$reset!$};
\draw (E2) to[out=110, in=70] (IDLE) node[above left =0.5cm and 0.1cm of E2 ] {$reset!$};

\end{tikzpicture}

	\caption{\label{fig:obsAsTA}An observer implemented using a timed automata. It detects if the process deadline (\mTPCproc) or sampling rate (\mTPCsamp) are not satisfied.  This observer implements the contract $\conSym_{C_1}^{1}$. For brevity, invariants are not shown. }
\end{figure}

Timed automata have been successfully used for fault diagnosis of industrial processes~\cite{Mhamdi2016}.
For the running example, the contract $\conSym_{C_1}^{1}$ specifies the timing requirements on the sampling time of the pulse counter (\mTPCproc) and the deadline for the processing data (\mTPCproc). The observer 
 is implemented using a timed automata~\cite{Alur:1994}, see Figure~\ref{fig:obsAsTA}.

\ignore{
\subsubsection{\acf{HA} as an observer}
Hybrid automata have been successfully used for modelling and monitoring of 
complex physical processes. They are more expressive than  \ac{TA}  as they do not restrict the rate of change of continuous variables  to one~\cite{Henzinger1996}.
For the running example, the contract $\conSym_{C_7}^{1}$ defines  a pressure leak, when the observed pressure ($p_{obs}(t)$)  is different from the expected pressure value by more than $5$~\%. The following observer is generated to monitor the pressure leak. 

\begin{figure}[hbtp]
	\centering
	\begin{tikzpicture}[->,>=stealth',shorten >=1pt,auto,
node distance=7cm,
semithick,scale=0.8, transform shape]
\tikzstyle{every state}=[rectangle,rounded corners,
minimum height = 1.5cm, text width=1.4cm, text centered, fill=blue!20,draw=none,text=black, draw,line width=0.3mm]
\tikzstyle{line} = [draw, -latex']

\node[state, fill=green!30, text width=3.4cm,
label={[shift={(0.0,-2)}]\small $ \mathbf{Process}$ }]
(PROC)  {$ {\dot{p}}_{exp} = k_1 \times  p_{exp}$};

\path[<-, dashed] (PROC.110) edge node[below, align=left, shift={(1.2,1)}] {
	 initial \\
	 $\begin{matrix}
	{p_{exp}(0)=k_0}{}
	\end{matrix}$
} ++(0cm,1cm);

\node[state, fill=red!20,
label={[shift={(0,0)}]$ $}, 
label={[shift={(0,-2)}]\small $\mathbf{Error}$   }]
(E1) [right of=PROC] {$\dot{t}=1$};



\draw (PROC) to[out=0, in=180] (E1) node[ above right =-0.7cm  and 0.2cm of PROC, text width =4cm] {$p_{obs}(t) < 0.95 \times p_{exp}(t)~or$ $p_{obs}(t) > 1.05 \times p_{exp}(t) $};


\end{tikzpicture}

	\caption{\label{fig:obsAsHA}An observer implemented using an \ac{HA}. It detects pressure leaks by comparing the sensor data ($p_{obs}$) against the expected data ($p_{exp}$). For brevity, invariants are not shown. }
\end{figure}
}
	

	\section{Implementation of the Cyber-infrastructure  }
	\label{sec:CIimplement}	
	
\label{sec:archImp}
	
In this section, we present implementation details of the 3-layered cyber infrastructure (see  Figure~\ref{fig:CILayers}) w.r.t. 
the assembly-line example (see Figure~\ref{fig:CIexample}). 
The implementation has an application, platform and physical layers as shown
in Figure~\ref{fig:CIarchImplementation}.

\subsection{Application layer}

The application layer accommodates the software components which describe the behaviour of an application. Earlier in Section~\ref{sec:componetOverview}, we described a component consisting of an interface, behaviours, contracts and observers. Furthermore, an application is specified as a network of components.

\subsubsection{\textbf{Component specifications.}} 
\begin{figure*}[hbtp]
	\centering
	\includegraphics[width=1.7\columnwidth]
	{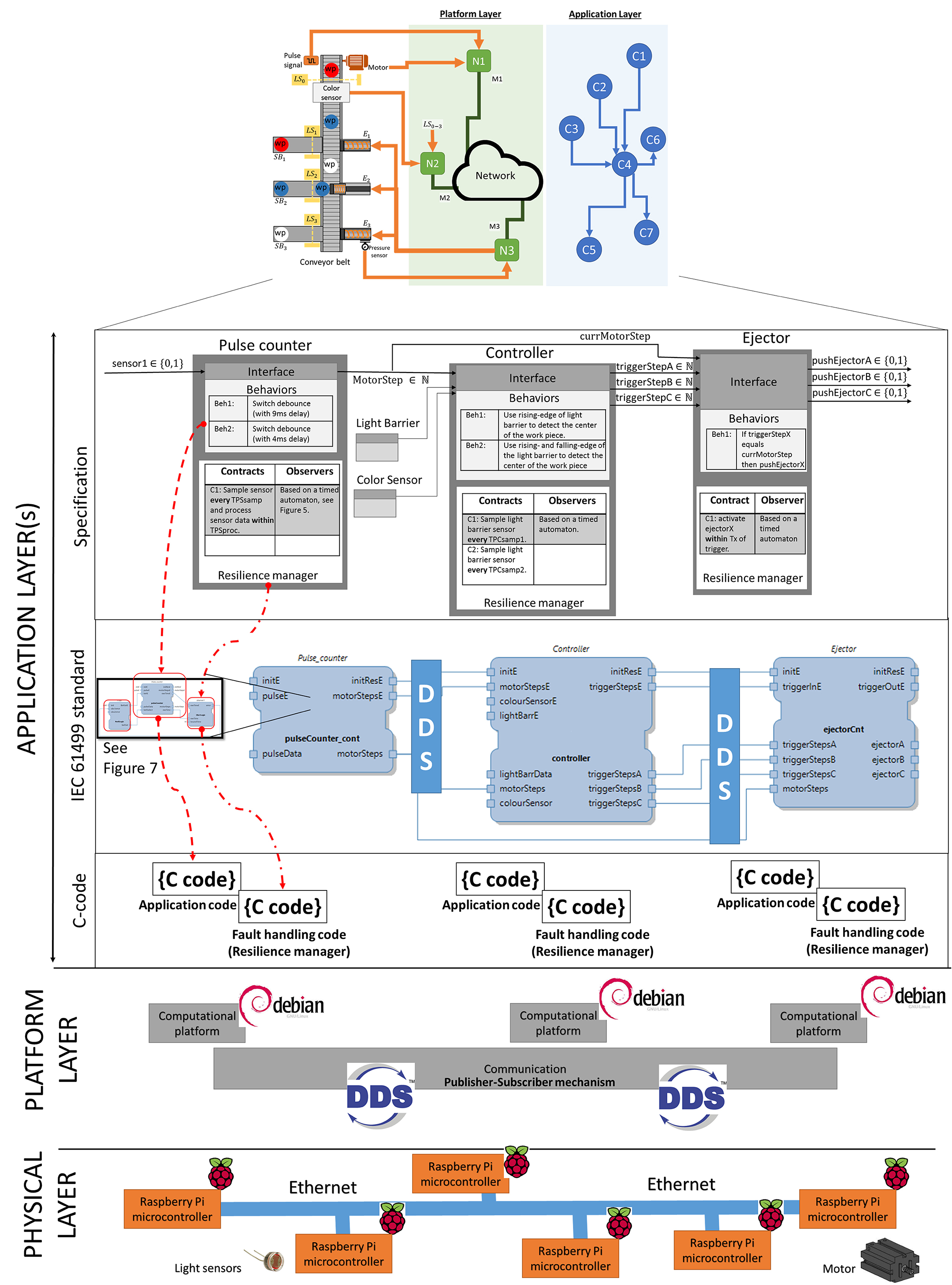}
	\caption{\label{fig:CIarchImplementation} An implementation of the proposed contract based approach for resilience. Only few components of the running example are shown due to brevity.    \vspace{-1.5cm}   }
\end{figure*}

An example specification with \codeFont{Pulse counter} (Component~$C_1$), 
\codeFont{Controller} (Component~$C_4$) and \codeFont{Ejector} (Component~$C_5$) is presented in Figure~\ref{fig:CIarchImplementation}. 
(a) Interface: The  \codeFont{Pulse counter} has one sensor input ($0$ or $1$) which is a rotary encoder  connected to a mechanical switch.
It has one output \codeFont{MotorStep} that represents the movement of the conveyor belt. For example, a work piece takes $7$ conveyor-belt steps to move from the first light barrier to the colour sensor.
(b) Behaviours: The component \codeFont{Pulse counter}  has two behaviours to address the noise from the mechanical switch. 
In \codeFont{Beh1}, a delay of $9$~ms is used when implementing the de-bounce functionality. In \codeFont{Beh2}, the safety margin is reduced such that the delay period is only $4$~ms.
(c) 
The  contract captures the requirement on the sampling frequency
(every \mTPCsamp) and the deadline on the processing time (within \mTPCproc). 
The corresponding observer for monitoring the contract at runtime is modelled using timed automata, see Figure~\ref{fig:obsAsTA}.
Finally, at runtime the resilience manager switches between behaviours  to satisfy the contract.  
Later in benchmarking, we demonstrate the response of the resilience manager when an observer fails.

\begin{figure*}[bhtp]
	\centering
	\includegraphics[width=1.5\columnwidth]
	{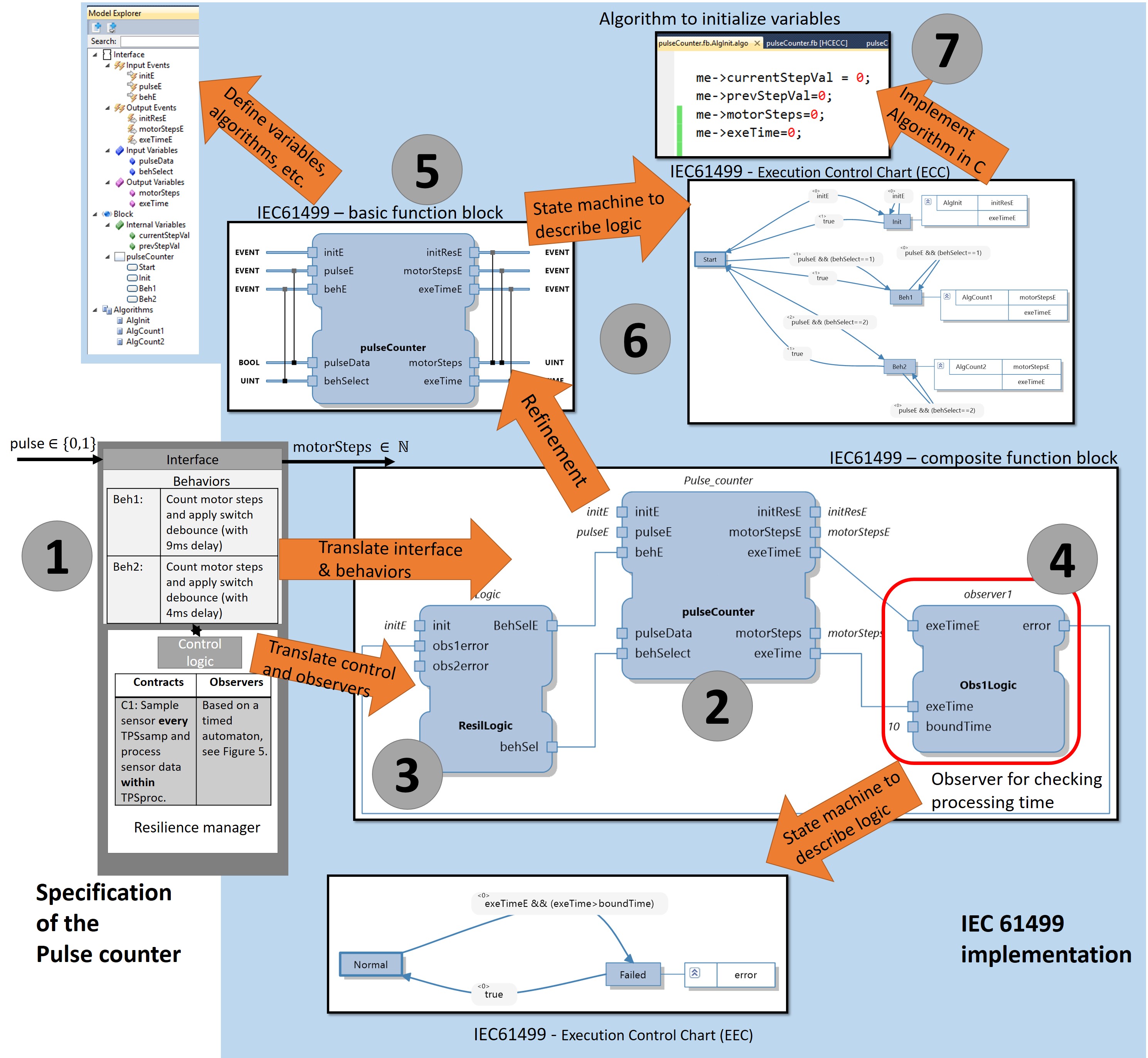}
	\caption{\label{fig:Spec2IEC} Translating the specification of the \codeFont{Pulse counter} component (interface, behaviours, observers and the resilience manager) to IEC~61499 standard (function blocks and control chart with states and algorithms). 
	\vspace{-1.5cm}}
\end{figure*}

\subsubsection{\textbf{Translating components to function blocks of IEC~61499 standard.}}

IEC 61499 framework represents a component-based solution for distributed industrial automation systems~\cite{Yoong2015}. An application is described as a network of \emph{function blocks}, see Figure~\ref{fig:Spec2IEC}.
Given the description of a component (see marker~1, in Figure~\ref{fig:Spec2IEC}), we present the translation to IEC 61499 standard.

 The component's interface and behaviours are mapped to a basic function block, see marker~2.
Due to the event-driven semantics of IEC61499, we need to associate input and outputs with events. E.g.,  \codeFont{pulse} is mapped to the input event \codeFont{pulseE} and the input variable \codeFont{pulseData}. When the event occurs the associated  data is updated internally, see inputs of the block near marker~5. 
The control logic of the function block is described using an Execution Control Chart (ECC). It receives  input events, and according to the current state, executes associated algorithms and emits events.
Marker~6 depicts an ECC with 4 states. They are \codeFont{Start}, \codeFont{Init}, \codeFont{Beh1} and \codeFont{Beh2}. 
When  the control reaches the \codeFont{Init} state, 
the associated algorithm \codeFont{AlgInit}  
initialises the internal and output variables (see marker~7). The IEC 61499 standard allows the behaviour to be described using C-language.

The component's resilience manager consists of a control logic  and a set of observers that monitor the contracts. 
 At runtime the resilience manager switches between behaviours (using the \codeFont{behSelect}) to satisfy the contracts. 
 An example observer based on timed automata is implemented using a timer, see marker~4. The ECC shows that the observer monitors the time between pulses (\codeFont{pulseE}) to ensure that the maximum time between two samples is satisfied (\codeFont{\mTPCsamp}) as input to the timer).
 For more details about the standard and the function blocks, see~\cite{Yoong2015}.
Later in the benchmarking section, we demonstrate the response of the resilience manager when an observer fails.

\subsubsection{\textbf{Translating IEC~61499 standard to executable C-code}}  Scheduling  of function blocks can be based on either an event-triggered or a cyclic-execution model. 
Event-triggered scheduling executes a function block when one of the input events are triggered.
 After execution of the block, the block may emit events which may trigger execution of another blocks. 
A queuing mechanism is used to address multiple events.  FORTE is one such runtime environment which is integrated with 4DIAC IDE~\cite{DIAC4tool,Zoitl2013}. Cyclic execution is an alternative scheduling model which resembles \ac{PLC} scan cycles. 
Here all function blocks are executed only once in each cycle. This execution model is supported by an off-the-shelf tool, called ISaGRAF~\cite{ISaGRAF}. 
Both tools depend on a complex runtime environment that is computationally intensive and cannot guarantee a deterministic and a \emph{deadlock-free} execution.  In contrast, a synchronous approach for the execution of function blocks has been developed~\cite{Yoong2015,Bhatti2011}. It does not require a runtime environment and provides a deterministic and deadlock free code.
The tool generated C-code  can be easily executed on a micro-controller. This provides flexibility for our implementation as we integrate other technologies such as \ac{DDS}. Figure~\ref{fig:CIarchImplementation} shows the generated C-codes for the function blocks.

\subsection{Platform layer}

The platform layer embodies computational and communicational platforms. 
In figure~\ref{fig:CIarchImplementation}, the generated C-code is executed on a Linux based computational platforms.  More specifically, we use Raspbian GNU/Linux~8.0 operating system (kernel version 4.9.35-v7)~\cite{Raspbian}.

Communication across the computational nodes is governed by the networking middle-ware, called, \acf{DDS}~\cite{Pardo2003,DDSrti}. DDS enables mechanisms that go beyond the classic publish-subscribe model. It can manage the interruptions when a publisher/subscriber  is temporarily or permanently unavailable. Furthermore,  it allows us to specify QoS parameters over the communication between a publisher and a subscriber. 
These parameters can be integrated into the resilience manager using contracts.  

\subsection{Physical layer}

All the sensors (light barriers, colour sensor, pulse switch) and actuators
 (motors, air compressor, ejectors) are part of the development kit from Fischertechnik~(product~\#536633).
Most of the computation is performed  by the microcontrollers. We have chosen Raspberry Pi 3 due to its flexibility. The processor is a quad core executing at 1.2~GHz  and has 1~GB RAM. A major drawback is that it does not support  analogue to digital converters.  
As a cost-effective solution, we used Arduino Pro Mini development board.
Ethernet is used for connecting all the Raspberry Pis.

	\section{An industrial cases study  }
	\label{sec:benchmarking}

\begin{figure}[hbtp]
	\centering
	\includegraphics[width=0.6\columnwidth]
	{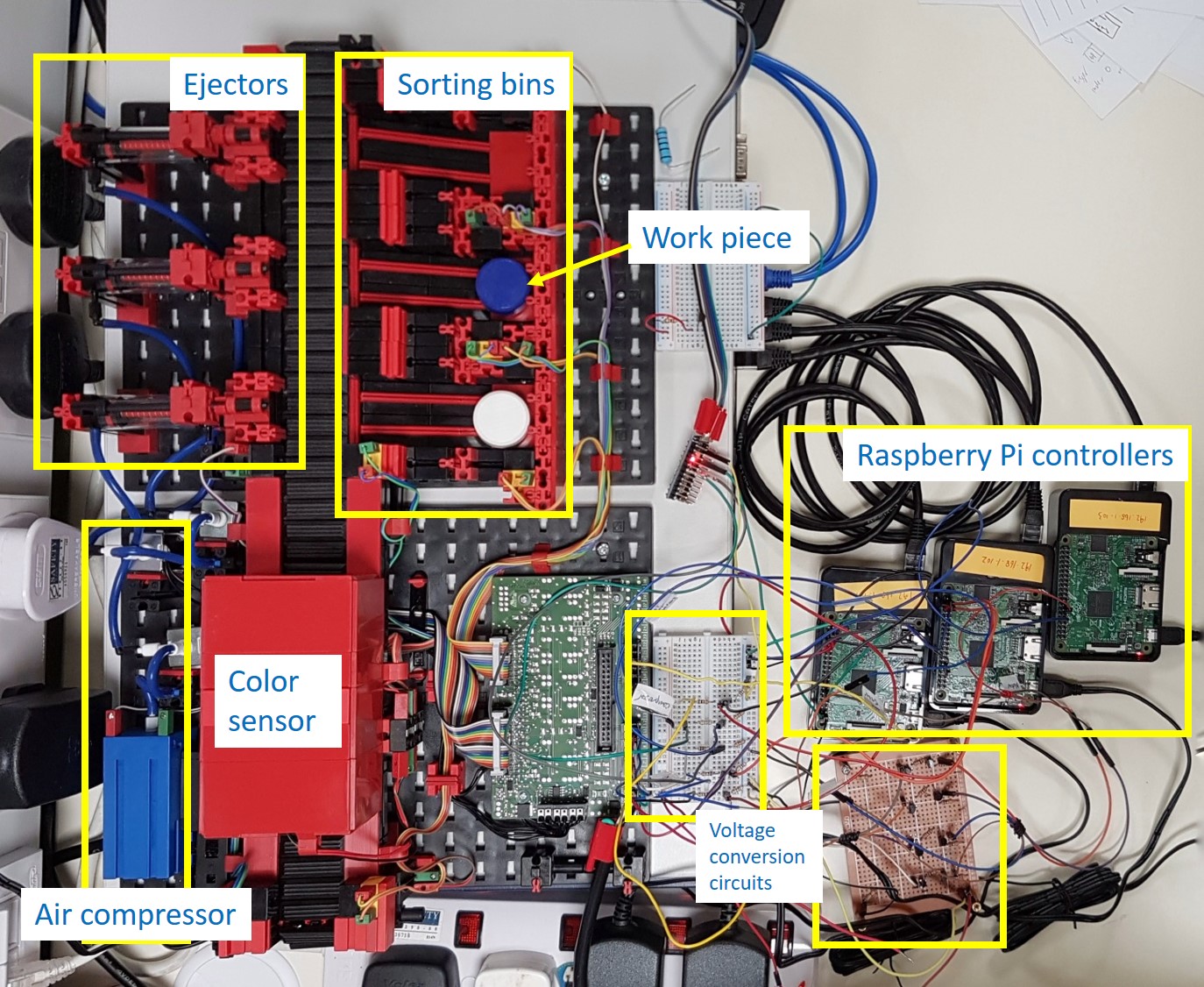}
	\caption{
		\label{fig:testbed} Evaluation tested reflecting the assembly line sorting application presented earlier in Figure~\ref{fig:CIexample}.   
\vspace{-1.5cm}	}
\end{figure}

In this section, we describe the benchmarking process used to validate the 
 proposed  contract-based approach. Figure~\ref{fig:testbed} shows an implementation of  the assembly line sorting application presented earlier in Figure~\ref{fig:CIexample}. Using the testbed, we implemented and deployed our resilient cyber-infrastructure and the contract-based approach.
 In the following, we present two experiments. First, we validate if  the end-to-end timing requirement  of the application is satisfied.
 Second, we present a scenario where a resilience manager of a component changes the component's behaviour when a fault is detected. We observe the fault-detection and recovery times.

\subsection{Experiment 1: validating end-to-end timing constraints}

\begin{figure}[hbtp]
	\includegraphics[width=1\columnwidth]
	{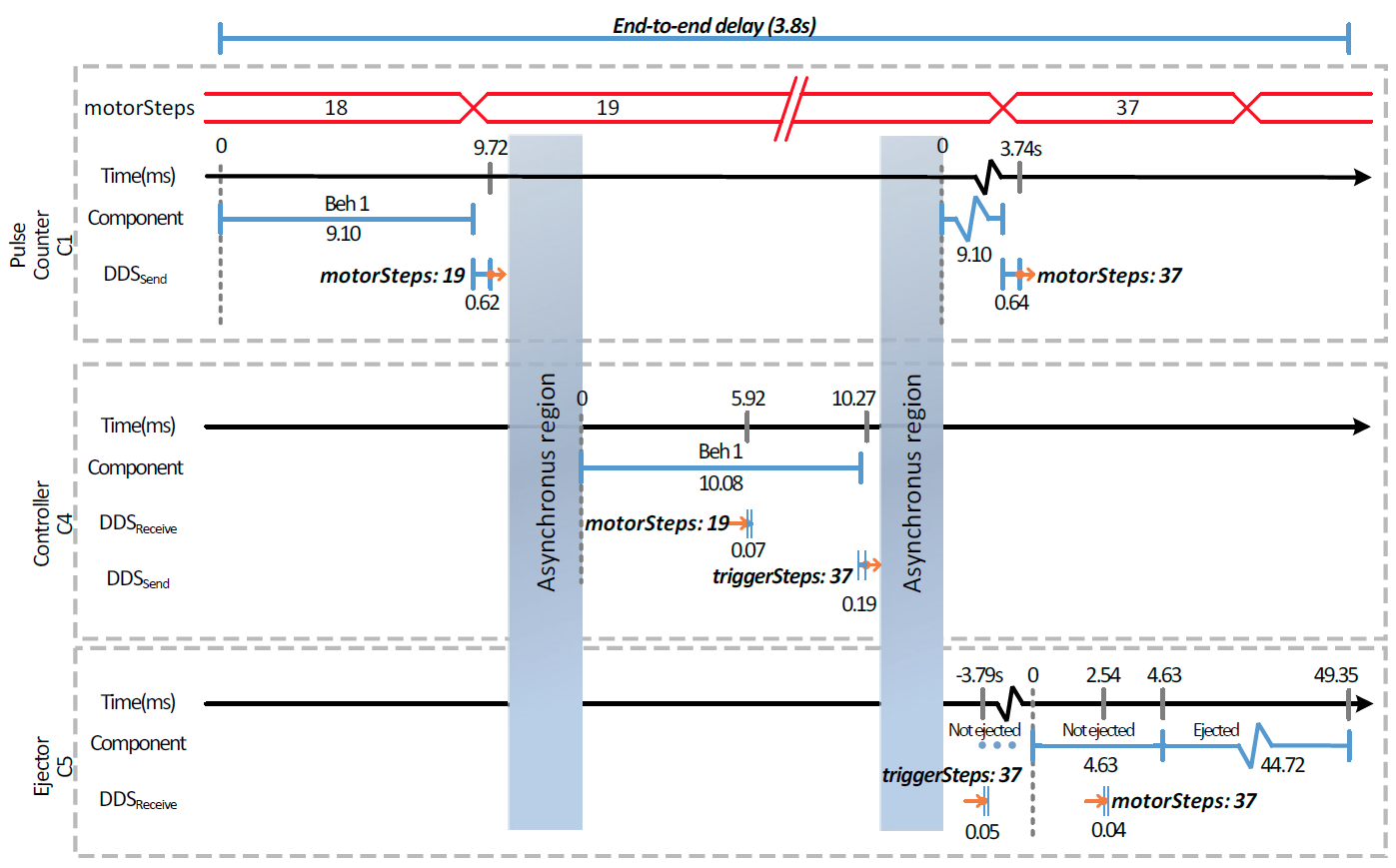}
	\caption{
		\label{fig:normalExe} Execution of the distributed system (without any faults).    
	}
\end{figure}

Earlier in Section~\ref{sec:appReq}, we presented that the end-to-end delay of the application should be less than~$4$ seconds. A failure to meet the deadline may cause a late activation of an ejector. This means a work piece is unable to reach its respective storage bin. In this experiment we validate the implementation by analysing an execution time sequence graph.  
Figure~\ref{fig:normalExe} presents the graph for the three components (\codeFont{Pulse counter}, \codeFont{Controller} and \codeFont{Ejector}) that are of interest. The three components are executing in parallel on three different micro controllers. 
We observe that the \codeFont{Pulse counter} executes its behaviour \codeFont{Beh~1} for a duration of $9.1$~ms. 
It computes the value of the new  \codeFont{MotorSteps} as~$19$. This information is then published by the DDS. The total duration to compute and send \codeFont{motorSteps} is $9.72$~ms. In parallel, \codeFont{Controller} receives the data and computes the new value for \codeFont{triggerSteps} as $37$.
 In parallel, \codeFont{Ejector} awaits  the value of \codeFont{triggerSteps} to match the value of current \codeFont{motorSteps} which is periodically sent by the \codeFont{Pulse counter}.
 The ejector is then activated to push a work piece from the assembly line. Finally, the measured end-to-end delay of the implementation is $3.8$ seconds which is less than the required $4$~seconds.

\subsection{Experiment 2: validating the response to a fault}

\begin{figure}[hbtp]
 	\includegraphics[width=1\columnwidth]
 	{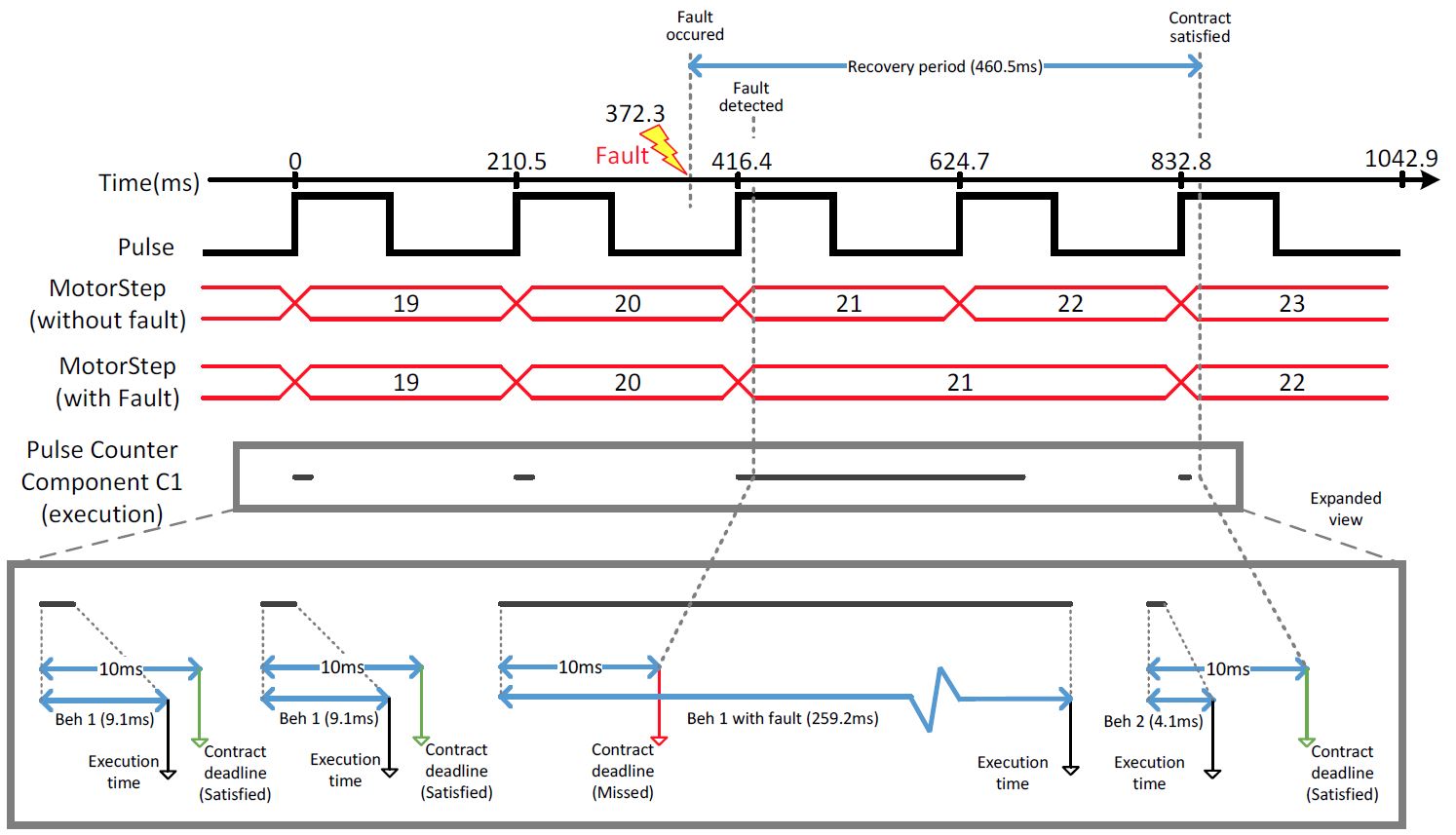}
 	\caption{\label{fig:faultDetection} Execution time sequence graph of the \codeFont{Pulse counter} \vspace{-2.5cm}}
 \end{figure}

For the \codeFont{Pulse counter} (Component $C_1$) presented in Figure~\ref{fig:normalExe},
the execution time sequence graph is presented in 
Figure~\ref{fig:faultDetection}.
The objective of the component is to increment the value of \codeFont{motorStep}
for every pulse, see the figure. The pulse signal is generated from a mechanical switch (the transient noise is not shown in the figure). 
 To address the noise, a simple denounce algorithm with waiting time of $9$~ms and $4$~ms is implemented by \codeFont{Beh~1} and \codeFont{Beh~2}, respectively. Furthermore, as explained earlier, the deadline for processing  the pulse sensor is $10$~ms.

 During the first two pulses ($0$ to $300$~ms) the component does not experience any faults. The contract (deadline of $10$~ms) is always satisfied because the execution time of \codeFont{Beh 1} is always less than $10$~ms.
 During the third pulse, a fault occurs in the computational platform which results in longer execution  of \codeFont{Beh~1}. 
 From the figure, we observe  the execution of \codeFont{Beh~1} to be $259.2$~ms due to the fault.
 This violates the contract after $10$~ms. 
 This is also when the architecture detects the fault. 
 The resilience manger decides to change the behaviour of the component from \codeFont{Beh~1}   to \codeFont{Beh~2}. However, we see the impact on the execution changing only in the fifth pulse.
 
 The \emph{recovery period}  for the component
 is the time from when \emph{fault occurred} to the point when the contract is once again satisfied. For our example, the recovery period is $460.5$~ms, see  Figure~\ref{fig:faultDetection}.
Due to the fault, the end-to-end deadline was not satisfied at the application layer. 
 This resulted in an incorrect  \codeFont{motorStep} value. 
 In the worst case, all work pieces that are on the conveyor belt when the fault occurred may have been sorted incorrectly. 
  Thus, the recovery period is equivalent to the end-to-end delay of the application which is approximately less than~$4$ seconds. 
  To reduce the application-level recovery time, we can communicate the missed \codeFont{motorStep} information to the \codeFont{Controller} which can adjust the \codeFont{triggerStep} for when the ejectors are to be activated.

	
	\section{Conclusions \& Future work }
	\label{sec:conclusions}
	
To enable a resilient cyber-infrastructure for Industry~4.0, we have presented a new contract-based methodology called CLAIR.
Applications are described as a set of modular components
that are distributed over a network. Contracts are used for
describing the component's interaction with other components (within and across layers).
Finally, the contract are monitored using runtime observers.
We detect failures (contract violation) and react (change of contracts) to the disturbances,  providing resiliency. 
Finally, using an industrial case study we have validated the  proposed architecture.

In future,  we plan to explore  efficient communication between resilience managers to reduce system-level recovery period.
Also, develop a multi-dimensional resilience metric to  evaluate resilience with respect to different performance indicators such as  safety, throughput, recovery time, etc.




	\bibliographystyle{ieeetr}
	\bibliography{references}



	
	


\end{document}